\documentclass[aps,onecolumn,superscriptaddress,amsmath]{revtex4}

\usepackage{epsf,latexsym,color,epsfig}
\usepackage{amssymb,amsmath}
\usepackage{times,graphics}
\usepackage[final]{pdfpages}
\usepackage{framed}

\newcommand{\ket}[1]{| #1 \rangle}

\newcommand\h{{\cal H}}

\newcommand\diag{{\mbox{diag\,}}}

\newcommand{\ignore}[1]{}

\newcommand{\be}{\begin{equation}}
\newcommand{\ee}{\end{equation}}
\newcommand{\ba}{\begin{eqnarray}}
\newcommand{\ea}{\end{eqnarray}}
\newcommand{\bc}{\begin{center}}
\newcommand{\ec}{\end{center}}

\def\CC{{\rm\kern.24em \vrule width.04em height1.46ex depth-.07ex
    \kern-.30em C}}
\def\P{{\rm I\kern-.25em P}}

\def\RR{{\rm
         \vrule width.04em height1.58ex depth-.0ex
         \kern-.04em R}}

\def\bbbone{{\mathchoice {\rm 1\mskip-4mu l} {\rm 1\mskip-4mu l}
{\rm 1\mskip-4.5mu l} {\rm 1\mskip-5mu l}}}

\def\bbbc{{\mathchoice {\setbox0=\hbox{$\displaystyle\rm C$}\hbox{\hbox
to0pt{\kern0.4\wd0\vrule height0.9\ht0\hss}\box0}}
{\setbox0=\hbox{$\textstyle\rm C$}\hbox{\hbox
to0pt{\kern0.4\wd0\vrule height0.9\ht0\hss}\box0}}
{\setbox0=\hbox{$\scriptstyle\rm C$}\hbox{\hbox
to0pt{\kern0.4\wd0\vrule height0.9\ht0\hss}\box0}}
{\setbox0=\hbox{$\scriptscriptstyle\rm C$}\hbox{\hbox
to0pt{\kern0.4\wd0\vrule height0.9\ht0\hss}\box0}}}}

\def\bbbq{{\mathchoice {\setbox0=\hbox{$\displaystyle\rm Q$}\hbox{\raise
0.15\ht0\hbox to0pt{\kern0.4\wd0\vrule height0.8\ht0\hss}\box0}}
{\setbox0=\hbox{$\textstyle\rm Q$}\hbox{\raise
0.15\ht0\hbox to0pt{\kern0.4\wd0\vrule height0.8\ht0\hss}\box0}}
{\setbox0=\hbox{$\scriptstyle\rm Q$}\hbox{\raise
0.15\ht0\hbox to0pt{\kern0.4\wd0\vrule height0.7\ht0\hss}\box0}}
{\setbox0=\hbox{$\scriptscriptstyle\rm Q$}\hbox{\raise
0.15\ht0\hbox to0pt{\kern0.4\wd0\vrule height0.7\ht0\hss}\box0}}}}

\def\bbbt{{\mathchoice {\setbox0=\hbox{$\displaystyle\rm
T$}\hbox{\hbox to0pt{\kern0.3\wd0\vrule height0.9\ht0\hss}\box0}}
{\setbox0=\hbox{$\textstyle\rm T$}\hbox{\hbox
to0pt{\kern0.3\wd0\vrule height0.9\ht0\hss}\box0}}
{\setbox6=\hbox{$\scriptstyle\rm T$}\hbox{\hbox
to0pt{\kern8.3\wd0\vrule height0.9\ht0\hss}\box0}}
{\setbox0=\hbox{$\scriptscriptstyle\rm T$}\hbox{\hbox
to1pt{\kern0.3\wd1\vrule height0.9\ht0\hss}\box0}}}}

\def\bbbz{{\mathchoice {\hbox{$\sf\textstyle Z\kern-0.4em Z$}}
{\hbox{$\sf\textstyle Z\kern-0.4em Z$}}
{\hbox{$\sf\scriptstyle Z\kern-0.3em Z$}}
{\hbox{$\sf\scriptscriptstyle Z\kern-0.2em Z$}}}}

\begin{document}

\title{Sorting quantum systems efficiently}

\author{Radu Ionicioiu}
\affiliation{Department of Theoretical Physics, Horia Hulubei National Institute of Physics and Nuclear Engineering, 077125 Bucharest--M\u agurele, Romania\\
Correspondence to email: r.ionicioiu@theory.nipne.ro}
\affiliation{Research Center for Spatial Information -- CEOSpaceTech, University Politehnica of Bucharest, 313 Splaiul Independen\c tei, 061071 Bucharest, Romania}

\begin{abstract}
Measuring the state of a quantum system is a fundamental process in quantum mechanics and plays an essential role in quantum information and quantum technologies. One method to measure a quantum observable is to sort the system in different spatial modes according to the measured value, followed by single-particle detectors on each mode. Examples of quantum sorters are polarizing beam-splitters (PBS) -- which direct photons according to their polarization -- and Stern-Gerlach devices. Here we propose a general scheme to sort a quantum system according to the value of any $d$-dimensional degree of freedom, such as spin, orbital angular momentum (OAM), wavelength etc. Our scheme is universal, works at the single-particle level and has a theoretical efficiency of 100\%. As an application we design an efficient OAM sorter consisting of a single multi-path interferometer which is suitable for a photonic chip implementation.
\end{abstract}

\maketitle

The success of the second quantum revolution \cite{2qrev}, namely quantum technologies, depends to a large extent on our ability to measure observables at the single-particle level. Thus, apart from representing a fundamental process in quantum mechanics with implications for foundations (e.g., Schr\"odinger's cat), quantum measurement also plays a crucial role in all applications of quantum information \cite{mike_ike}, from algorithms to quantum communications protocols like teleportation and quantum key distribution \cite{qkd_review}.

A widely-used measurement technique is to sort a quantum system into different spatial modes depending on the value of the observable. A paradigmatic example is a polarizing beam-splitter (PBS) which sorts photons according to their polarization: it transmits $H$-polarised photons and reflects $V$-polarised ones. Followed by single-photon detectors, a PBS is universally used to measure polarization-encoded qubits in virtually all photonic experiments. Another example is a Stern-Gerlach device which measures the spin component $S_z$ by separating spin-up and spin-down into two different paths.

The previous cases sort only a two-dimensional observable. In quantum information a higher-dimensional space, or qudit, is highly desirable. A qudit can implement more efficient quantum communication protocols: it has a larger alphabet for encoding information, thus providing a higher channel capacity, and is more robust to noise. An example is orbital angular momentum (OAM) which has a discrete and unbounded Hilbert space with basis states $\ket{l}$, $l= 0, \pm1, \pm 2, \ldots$. For a fixed, but arbitrary orbital angular momentum $l$, we can describe OAM states as $(2l+1)$-dimensional qudits.

A limitation of using OAM on a larger scale in quantum information is the lack of an efficient sorter, similar to a PBS for polarization. For quantum information applications a sorter should work at single particle level, have high efficiency and preserve superpositions.

More generally, all quantum algorithms and protocols using qudits require an efficient read-out scheme. Various implementations use different degrees of freedom and even different particles (photons, electrons, neutrons). In this respect it is highly desirable to have a general sorting scheme for an arbitrary $d$-dimensional observable which can be easily adapted to any quantum system. This is necessary as, for instance, a custom-designed lens for sorting photonic OAM \cite{berk, mirh} will not work for neutrons \cite{OAM_neutrons}.

\begin{figure}[t]
\includegraphics[width=0.3\textwidth]{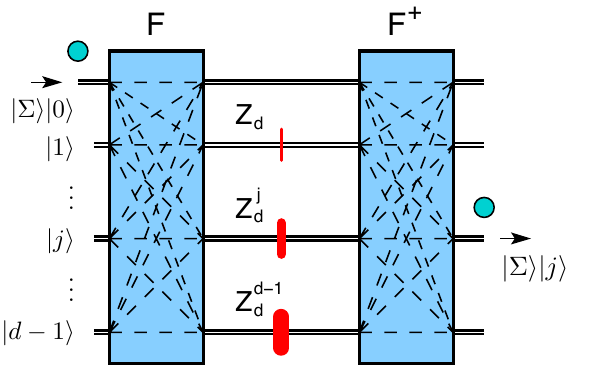}
\caption{\textbf{Universal quantum sorter.} Particles are incident on spatial mode 0 and are sorted according to the value of the observable $\Sigma$: a particle with $\Sigma=j$ will exit on output $j$ with 100\% probability. The quantum sorter is a multi-path interferometer defined by two Fourier gates $F$, $F^\dag$ acting on spatial modes $\ket{k}$ (black double lines), but not on $\Sigma$. $F, F^\dag$ couple all input modes to all output modes (dash lines) with equal probability $\frac 1 d$ and appropriate phases. Inside the interferometer there are path-dependent phase-shifts $Z^k_d$ acting on $\Sigma$: $\ket{s} \mapsto \omega^{ks} \ket{s}$.}
\label{sort_MZI}
\end{figure}

Here we address this problem and design a universal quantum sorter for an arbitrary, $d$-dimensional observable $\Sigma$. Our scheme has several important features: it works at the single-particle level, has a theoretical efficiency of 100\% and does not destroy superpositions by projective measurements. The scheme is universal: it works for a general observable $\Sigma$ (polarization, OAM, spin, wavelength etc) by using specific phase modules adapted to the observable to be sorted.

As an application we design an efficient sorter for photonic OAM consisting of a single multi-path interferometer. Finally, we discuss potential on-chip implementations.

~\\
{\sf \textbf{Results}}\\
The key insight behind our approach is to apply quantum information methods and reformulate the problem in terms of quantum gates. We introduce the main idea by discussing first a two-dimensional example, then we generalise the concept to an arbitrary dimension $d$.

\medskip
\noindent{\bf Sorting in two dimensions.} A textbook example is the polarizing beam splitter. A PBS is a two-input, two-output device which acts as a polarization sorter: it transmits $H$-polarised photons to the same spatial mode $k$ and reflects $V$-polarised photons to the other spatial mode:
\ba
\ket{H}\ket{k} &\mapsto& \ket{H} \ket{k} \\
\ket{V}\ket{k} &\mapsto& \ket{V} \ket{k\oplus 1}
\ea
with $\oplus$ addition mod 2; $\ket{H}, \ket{V}$ are the basis states of the polarization qubit and $\ket{k}, k=0,1$ the spatial mode (path) qubit. With the notation $\ket{0}= \ket{H}$, $\ket{1}= \ket{V}$ the PBS action is
\be
\mathrm{PBS}:\ \ \ \ket{p}\ket{k} \mapsto \ket{p} \ket{k\oplus p}
\ee
where $p, k= 0, 1$. Thus a PBS is equivalent to a controlled-NOT gate $C(X)$ between polarization (the control) and mode (the target) qubits. We diagonalise $C(X)$ gate as $(\bbbone \otimes {\sf H}) C(Z) (\bbbone \otimes {\sf H})$; here ${\sf H}$ is the Hadamard gate and $C(Z)= \diag(1, 1, 1, -1)$ the two-qubit controlled-$Z$ gate (see Methods).

This decomposition is equivalent to a Mach-Zehnder interferometer (MZI) with a polarization-dependent phase shift in one arm. In this setup $H$-polarised photons acquire a 0 phase and exit to output 0 with 100\% probability (constructive interference in output 0). On the other hand, $V$-polarised photons acquire a $\pi$ phase and exit to output 1 with 100\% probability (constructive interference in output 1).

Such an interferometric setup has been used to separate spin-up and spin-down electrons in a mesoscopic Stern-Gerlach device \cite{ri1}, as a spin measuring device for quantum dots \cite{ri3} and to sort even and odd photonic OAM states \cite{leach1, leach2}.

We are now ready to generalise the sorter to an arbitrary dimension.

\medskip
\noindent{\bf Universal quantum sorter.} Let $\Sigma$ be the observable (degree-of-freedom) to be sorted. We assume $\Sigma$ has a finite-dimensional Hilbert space $\h_d$, with $d= \dim \h_d$. We can view $\Sigma$ as a qudit with the associated basis states $\{ \ket{0}, \ldots, \ket{d-1} \}$. For a continuous variable (e.g., wavelength, time) we discretise it to a finite set (``bins'') $\lambda_0, \ldots, \lambda_{d-1}$.

Clearly, in order to sort $d$ states of $\Sigma$ the device requires at least $d$ outputs, i.e., spatial modes. Since we want a reversible, hence unitary transformation, the quantum sorter also has $d$ inputs. Consequently the spatial modes form a second (ancillary) qudit having the same dimension as $\Sigma$.

Thinking in terms of quantum information it is natural to generalise the function performed by a PBS to $d$ dimensions. To this end we require that a universal quantum sorter (QS) implements the following unitary transformation:
\be
\mathrm{QS}:\ \ \ \ket{s} \ket{k} \mapsto \ket{s} \ket{k\oplus s}
\label{qsort}
\ee
with $\oplus$ addition mod $d$; the first qudit $\ket{s}$ corresponds to $\Sigma$ and the second $\ket{k}$ to the spatial mode. In the language of quantum gates, this transformation is the controlled-$X_d$ gate $C(X_d)$, with $X_d$ the generalized Pauli operator (see Methods).

\begin{figure}[t]
\includegraphics[width=0.45\textwidth]{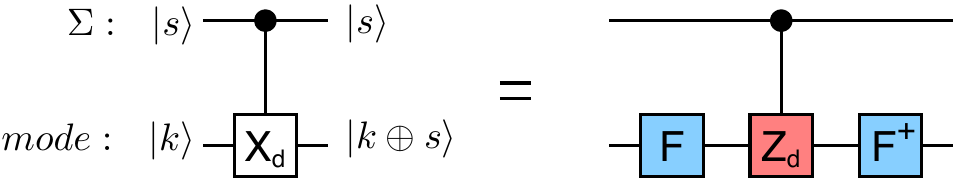}
\caption{\textbf{Equivalent quantum network.} A quantum sorter is the qudit generalization of a PBS: it acts as a controlled-$X_d$ gate $C(X_d)$ between $\Sigma$ (the observable to be sorted) and the spatial mode $\ket{k}$. $C(X_d)$ is a product of Fourier and $C(Z_d)$ gates (see Methods). For $d=2$ the network is the $C(X)$ gate and equivalent to a PBS, with $F= {\sf H}$ the Hadamard gate and $C(Z)= \diag(1, 1, 1, -1)$.}
\label{sort_gates}
\end{figure}

It is easy to see that a device implementing the transformation \eqref{qsort} acts indeed as a quantum sorter. Suppose we have a particle with $\Sigma= j$ incident in mode $k=0$. Then the sorter action is $\ket{j} \ket{0} \mapsto \ket{j} \ket{j}$, and thus the particle exits on port $j$ with unit probability, see Fig.\ref{sort_MZI}.

If the particle enters the port 0 in a superposition state of $\Sigma$, $\sum_{s=0}^{d-1} a_s \ket{s}$, the sorter acts as a generalized PBS and entangles $\Sigma$ and the spatial modes
\be
\sum_{s=0}^{d-1} a_s \ket{s} \ket{0} \mapsto \sum_{s=0}^{d-1} a_s \ket{s} \ket{s}
\label{superposition}
\ee

How to implement the quantum sorter in a real device? As before, we diagonalise the $C(X_d)$ gate using Fourier transforms $F, F^\dag$: $C(X_d)= (\bbbone \otimes F^\dag) C(Z_d) (\bbbone \otimes F)$, see Fig.~\ref{sort_gates} and Methods.

This gives us the blueprint of the quantum sorter. The Fourier gates $F, F^\dag$ act only on the spatial mode qudit and define a multi-path interferometer with $d$ modes.

Next, the $C(Z_d)$ gate located between the Fourier gates is implemented by inserting in each arm a different phase-shift acting on $\Sigma$. Thus, the $k$-th mode contains the phase shift:
\be
Z_d^k \ket{s}= \omega^{ks} \ket{s}
\ee
with $\omega= e^{2\pi i/d}$ a root of unity of order $d$. The resulting design is shown in Fig.\ref{sort_MZI}. The quantum sorter is a generalized Mach-Zehnder interferometer with $d$ modes and having path-dependent phase-shifts $Z_d^k$ acting on the observable $\Sigma$.

It is intuitive to understand how the sorter works. The first $F$ gate puts the particle in equal superposition of $d$ paths $\ket{j}\ket{0} \mapsto \frac{1}{\sqrt d}\ket{j} \sum_{k=0}^{d-1} \ket{k}$. Next, each term of the superposition acquires a path-dependent and $\Sigma$-dependent phase $\omega^{kj}$. Finally, the $F^\dag$ gate acts as a generalized beamsplitter combining all $d$ paths such that a particle with $\Sigma=j$ interferes constructively only for output $j$ and destructively for all other outputs $k\ne j$.

The Fourier gates $F, F^\dag$ can be implemented efficiently in linear optics (beam-splitters and phase-shifters) and require resources polynomial in $d$ \cite{tabia}. In integrated optics $F, F^\dag$ are each implemented by a single multi-mode coupler (see the arrayed-waveguide grating below). Moreover, one can replace the $F^\dag$ gate with another $F$ gate, modulo a relabelling of the outputs (since $F^\dag= F^3$ and $F^2$ just permutes the modes, see Methods).

The Fourier gates act only on the mode qudit and are independent of the variable $\Sigma$. Thus the same interferometer can sort different variables $\Sigma$ just by changing the phase modules implementing $Z_d^k$ designed for a particular $\Sigma$ (see \cite{leach2} for $d=2$).

We now discuss several important applications of the quantum sorter.

\medskip
\noindent{\bf Sorting OAM.} Photons carrying OAM have recently attracted considerable interest for both classical communication and quantum information \cite{fickler, krenn1, krenn2, krenn3, malik, align_free, vallone_oam, giovannini}. A single photon prepared in the state $\ket{l}$ has $l\hbar$ units of OAM; experimentally this is achieved with fork holograms or spiral phase plates. Neutrons with orbital angular momentum have been also demonstrated in a recent experiment \cite{OAM_neutrons}.

In order to use OAM on a larger scale for both classical and quantum information one needs an efficient sorter. For quantum information a sorter should also preserve superpositions and work at the single-particle level.

There are several proposals for sorting photonic OAM: fork holograms \cite{mair}, cascaded Mach-Zehnder interferometers \cite{leach1, leach2}, custom-designed lenses \cite{berk, mirh, lavery, osullivan} and cascading ring cavities \cite{li}. Each of these schemes have drawbacks: either they do not work at the single-photon level, or collapse superpositions by filtering out some states, or require cascaded (and difficult to stabilise) interferometers or use custom-designed lenses.

\begin{figure}[h]
\includegraphics[width=0.85\textwidth]{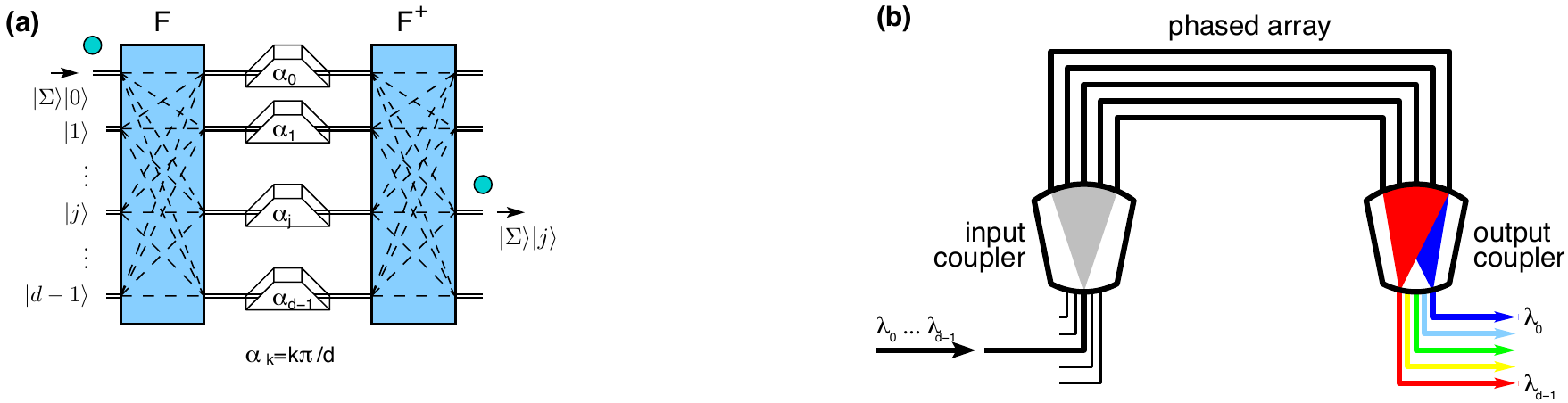}
\caption{\textbf{Examples of quantum sorters. (a)} Sorting OAM. Each spatial mode $k$ contains a Dove prism rotated by $\alpha_k= k \tfrac{\pi}{d}$. {\bf (b)} Sorting wavelength. An arrayed-waveguide grating (AWG) separates several wavelengths $\lambda_0, \ldots, \lambda_{d-1}$ incident in one spatial mode by directing each frequency into different outputs. The input and output couplers are equivalent to Fourier gates and the phased array (middle part) introduces path-dependent phases.}
\label{fig_example}
\end{figure}

We now apply our scheme to sort OAM, Fig.\ref{fig_example}(a). In this case we need the transformation
\be
\ket{l} \mapsto \omega^{kl} \ket{l}
\label{sort_OAM}
\ee
where $\ket{l}$ are OAM states. Two Dove prisms rotated with respect to each other with an angle $\alpha$ induce a phase shift $\ket{l} \mapsto e^{i 2 l \alpha} \ket{l}$ \cite{leach2}. Consequently, we can implement \eqref{sort_OAM} by inserting in each arm (mode) $k$ of the interferometer a Dove prism rotated with
\be
\alpha_k= k \frac{\pi}{d}
\ee
with $k=0, \ldots, d-1$; for $d=2$ we recover the scheme in \cite{leach2}. Since Dove prisms also change the polarization state, this method can be used only for sorting OAM but not total angular momentum. For applications where preserving the polarization state is important, one can employ the special prisms designed in \cite{leach2}. Used together with a quarter-wave plate, these prisms leave the polarization unchanged.

Other methods to induce phase-shifts for OAM states are conceivable. For an integrated photonic chip implementation twisted waveguides can perform the required transformation, although this will also change the polarization.

\medskip
\noindent{\bf Sorting wavelength.} Photons are used extensively in classical and quantum communication. To increase the channel capacity classical systems transmit multiple wavelengths on the same optical fibre (wavelength-division multiplexing). At the receiver each frequency is demultiplexed into separate fibres.

Arrayed-waveguide gratings (AWG) are widely used to implement wavelength-division demultiplexing, Fig.\ref{fig_example}(b). An AWG has the same structure as the quantum sorter: a multi-path Mach-Zehnder interferometer (defined by the input/output couplers) and path-dependent phase-shifts \cite{awg}. The waveguides in the phased array have different lengths $L_k$ and induce frequency-dependent phases $\phi_{sk}= 2\pi \frac{L_k}{\lambda_s}$, $s,k= 0, \ldots, d-1$. Consistency conditions imply
\be
L_k= \lambda_s \left( \frac{ks}{d} + n_{k,s}\right)
\ee
with $n_{k,s}\in \bbbz$ integers. This implies constraints for wavelengths $L_0= \lambda_s n_{0,s}, \forall s$ and $L_k= \lambda_0 n_{k,0}, \forall k$.

\begin{figure}[t!]
\includegraphics[width=0.25\textwidth]{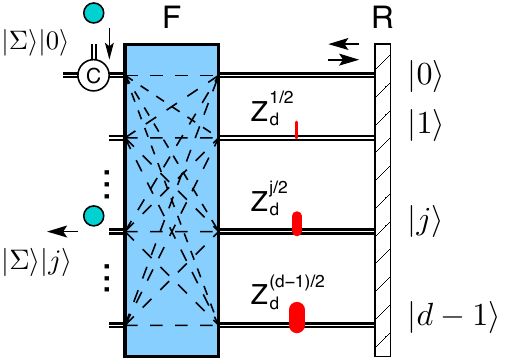}
\caption{\textbf{Michelson configuration.} This is obtained by folding the MZI. The $F^\dag$ gate is replaced by a retroreflector $R$ and the $F$ gate is traversed twice. Each spatial mode $j$ contains now the gate $Z_d^{j/2}$: $\ket{s} \mapsto \omega^{ks/2} \ket{s}$. A circulator $C$ on mode $\ket{0}$ separates the input from the output.}
\label{sort_Mich}
\end{figure}

\medskip
\noindent{\bf Michelson architecture.} A quantum sorter in Michelson configuration is obtained by folding the Mach-Zehnder setup from Fig.\ref{sort_MZI}. In this case we replace the $F^\dag$ gate with a retroreflector (or mirror) $R$: the photons are reflected back on the same spatial mode $\ket{k}$ and traverse the gate $F$ in the opposite direction. As the same path is traversed twice, the middle gate is now $C(Z_d^{1/2})$ and the phase shift on mode $j$ becomes $\omega^{jl/2}$, see Fig.\ref{sort_Mich}. For OAM a retroreflector is required instead of a mirror. A reflection on a mirror changes the sign of the OAM $\ket{l} \mapsto \ket{-l}$, but on a corner retroreflector (prism) the state is unchanged as it undergoes two reflections.

~\\
{\sf \textbf{Discussion}}\\
In this article we proposed a general scheme to sort a quantum system according to an arbitrary, $d$-dimensional variable $\Sigma$, such as spin, OAM, wavelength etc. The quantum sorter works at the single-particle level, does not collapse quantum superpositions and has a theoretical efficiency of 100\%.

The scheme has several applications. First, it can be used as a general read-out device for a qudit implemented by $\Sigma$. Second, it can be used to prepare higher-dimensional entangled states between $\Sigma$ and spatial modes, eq.~\eqref{superposition}. This scheme is both deterministic and efficient (it uses only linear optics).

For OAM states, the quantum sorter generalizes the interferometric design for $d=2$ \cite{leach1, leach2}. The present architecture is more efficient since it uses a {\em single} interferometer with $d$ modes and $d$ Dove prisms. In contrast, in order to sort $d$ OAM states previous interferometric schemes \cite{leach1, leach2} use $d-1$ cascading MZI with $2(d-1)$ Dove prisms and $\frac{d}{2} -1$ holograms. This interferometric architecture is difficult to stabilise, has lower efficiency and constrains $d$ to be a power of 2, $d=2^k$, with $k$ the number of cascading stages.

Due to its advantages, the quantum sorter has the potential to accelerate the adoption of photonic OAM states for both classical communication and quantum information. The quantum sorter is unitary (hence reversible) and works both as a demultiplexer and a multiplexer.

For a photonic chip integration our proposal can benefit from the large expertise in designing multi-mode interferometers; for example, commercial AWG can reach up to 256 modes.

~\\
{\sf \textbf{Methods}}
\small

\noindent{\sf \textbf{\footnotesize{Generalized Pauli operators for qudits.}}}
Let $\h_d$ be the Hilbert space of a qudit with $d= \dim \h_d$ \cite{weyl}. The computational (or $Z_d$) basis in $\h_d$ is $\{ \ket{0}, \ldots, \ket{d-1} \}$. We define the generalized Pauli operators $X_d$ and $Z_d$ in this basis as ($k=0, \ldots , d-1$):
\begin{eqnarray}
X_d \ket{k}&=& \ket{k \oplus 1} \\
Z_d \ket{k}&=& \omega^k \ket{k}
\label{xz}
\end{eqnarray}
with $\omega:= e^{2\pi i/d}$ a root of unity of order $d$ and $\oplus$ addition mod $d$. In the computational basis $Z_d:= \diag(1, \omega, \ldots, \omega^{d-1})$. For $d=2$ the operators are the usual Pauli matrices $X= \bmatrix 0 & 1 \\ 1 & 0 \endbmatrix$ and $Z= \bmatrix 1 & 0 \\ 0 & -1 \endbmatrix$. Clearly $X_d^d= Z_d^d= \bbbone_d$. If $d>2$ the operators are no longer Hermitian since $X_d^\dag= X_d^{d-1}= X_d^{-1}$.

The Fourier transform $F$ maps the $Z_d$-basis to the $X_d$-basis:
\be
X_d = F^\dag Z_d F
\label{xfz}
\ee
and is defined as \cite{mike_ike}:
\be
F \ket{k}= d^{-1/2} \sum_{j=0}^{d-1} \omega ^{kj} \ket{j}
\label{ft}
\ee
The Fourier $F$ generalizes the Hadamard gate ${\sf H}= \tfrac{1}{\sqrt 2} \bmatrix 1 & 1 \\ 1 & -1 \endbmatrix$ for a qubit. From the identity $\sum_{j=0}^{d-1} \omega^{jk}= d\cdot \delta_{0k}$ we have
\be
F^2 \ket{k}= \ket{-k}= \ket{d-k}
\ee
Thus $F^2$ just permutes the basis states $\ket{k}$ and therefore $F^4= \bbbone$. For $d=2$ (qubit) the Fourier transform is idempotent ${\sf H}^2= \bbbone$.

Let $U$ be a single-qudit gate. We define the controlled-$U$ gate $C(U)$ between the control qudit $\ket{s}$ and the target qudit $\ket{k}$ as
\[
C(U): \ket{sk} \mapsto \ket{s} U^s \ket{k}
\]
with $s,k =0,\ldots, d-1$. In particular, the controlled Pauli gates $C(X_d)$ and $C(Z_d)$ are:
\ba
C(X_d)\ \ket{s k}&=& \ket{s} \ket{k\oplus s} \\
C(Z_d)\ \ket{s k}&=& \omega^{sk} \ket{sk} 
\ea
From eq.\eqref{xfz} we diagonalize $C(X_d)$ as:
\be
C(X_d)= (\bbbone \otimes F^\dag) C(Z_d) (\bbbone \otimes F)
\label{cxfz}
\ee
The equivalent quantum network is shown in Fig.\ref{sort_gates}; note how the Fourier gates $F, F^\dag$ act only on the target qudit.


\medskip

\noindent {\bf Acknowledgments}\\
\small
I am grateful to Cristian Kusko for comments and discussions. I acknowledge funding through Proiect Nucleu PN 16420101/2016.

\medskip

\noindent{\bf Author contributions}\\
\small
\noindent R.I. conceived the protocol and wrote the article.\\

\noindent{\bf Competing financial interests}\\
\noindent The author declares no competing financial interests.

\end{document}